\documentclass[onecolumn, preprintnumbers, showpacs, nofootinbib, aps]{revtex4-1}
\usepackage{graphicx}

\newcommand{\bi}{\bibitem}
\newcommand{\be}{\begin{eqnarray}}
\newcommand{\ee}{\end{eqnarray}}
\newcommand{\rar}{\rightarrow}

\begin{document}

\title{Shape and position of the shadow 
in the $\delta = 2$ Tomimatsu-Sato space-time}

\author{Cosimo Bambi}
\email{cosimo.bambi@ipmu.jp}

\author{Naoki Yoshida}
\email{naoki.yoshida@ipmu.jp}

\affiliation{Institute for the Physics and Mathematics of the Universe, 
The University of Tokyo, Kashiwa, Chiba 277-8583, Japan}

\date{\today}

\preprint{IPMU10-0061}

\begin{abstract}
Within 5-10 years, very long baseline interferometry facilities
will be able to observe the ``shadow'' of super-massive
black hole candidates. This will allow, for the first time, to 
test gravity in the strong field regime. In this paper, we study
numerically the photon orbits in the $\delta = 2$ Tomimatsu-Sato 
space-time. The $\delta = 2$ Tomimatsu-Sato space-time is 
a stationary, axisymmetric, and asymptotically flat exact solution 
of the vacuum Einstein equations. We compare the associated 
shadow with the one of Kerr black holes. The shape of the 
shadow in the $\delta = 2$ Tomimatsu-Sato space-time is oblate
and the difference between the two axes can be as high as 6\% 
when viewed on the equatorial plane. We argue that future space 
sub-mm interferometers (e.g. VSOP-3) may distinguish the two cases, 
and thus are able to test the Cosmic Censorship Conjecture.
\end{abstract}


\maketitle


\section{Introduction}

Today gravity is relatively well tested in the weak field limit, 
while little or nothing is known when it becomes strong~\cite{will}.
One of the most outstanding issues is the actual nature of the 
final product of the gravitational collapse. Currently
there are many known ``black hole candidates'', 
but we do not know if these
objects are really the black holes predicted by general relativity
and, more in general, if they have an event horizon. At the
observational level, the situation may however change soon: the
capability of very long baseline interferometry (VLBI) has
improved significantly at short wavelength and it is now widely
believed that within 5-10 years it will be possible to observe
the direct image of the accretion flow around a black hole with
a resolution comparable to its event horizon~\cite{doe1, doe2}.
Such observations will allow to test gravity in the strong field
regime and investigate the nature of our black hole candidates.

One of the main targets of these VLBI experiments is the observation
of the black hole ``shadow''~\cite{bardeen, melia, rohta}, i.e. a dark 
area over a bright background. The exact shape and position of the
shadow directly depend on the metric of the space-time very close
to the massive object, being determined by the innermost photon
orbits. In light of this, it is important
to understand how to use future observations to test general 
relativity and what questions can be addressed. For example,
in~\cite{sha1, sha2}, it is shown that the shadow can be used to
test the Kerr bound $|a_*| \le 1$, where $a_*$ is the dimensionless
spin parameter. In~\cite{psaltis}, the authors suggest the test
of the no-hair theorem. In this paper we study the shadow 
in the $\delta = 2$ Tomimatsu-Sato (TS2) space-time~\cite{tom1,tom2}, 
which is {\it not} a black hole solution. We argue that future sub-mm
interferometers will be able to measure the shape of the
shadow with an accuracy at the level of 1\% and thus may 
distinguish the Kerr from the TS2 space-time.

TS2 is a stationary, axisymmetric, and 
asymptotically flat exact solution of
the vacuum Einstein equations. The Kerr metric describes the
gravitational field produced by a spinning mass, while the TS2 one 
describes the gravitational field of a spinning and deformed mass.
Despite such appealing features, the solution violates the Cosmic
Censorship Conjecture~\cite{penrose}, 
which precludes the formation of naked singularities in the 
Cauchy development starting from regular initial data. In 4D 
general relativity, the Kerr space-time is the only stationary
and asymptotically flat solution with a regular event 
horizon~\cite{carter, robinson}. In presence of naked singularities,
there is no uniqueness theorem and the Tomimatsu-Sato space-times
are just the simplest extension of the Kerr solution.
In general, one can indeed expect that the massive object is
described by an infinite number of free parameters~\cite{m-n}.
However, it turns out that the introduction of a larger number
of parameters do not change significantly the orbits of the 
space-time: basically they represent corrections to higher order 
multipole moments. So,
as a first approximation, deviations from the Kerr metric can be
studied by considering the Tomimatsu-Sato space-times and here,
for the sake of simplicity, we discuss the TS2 solution, even if 
this does not mean that TS2 is the only alternative to the Kerr 
space-time when the Cosmic Censorship Conjecture is relaxed. It 
is however remarkable that TS2 can be obtained from the 
Neugebauer-Kramer solution representing a superposition of two 
Kerr black holes, in the limit in which the centers of the two black 
holes coincide. This fact has been interpreted in~\cite{kodama}
as an indication that TS2 can be a good candidate for the final 
stage of the gravitational collapse.

The Cosmic Censorship Conjecture is motivated by the fact that 
space-times with naked singularities present several kinds of 
pathologies.
In the case of the Tomimatsu-Sato space-times, the most suspicious 
property is the existence of closed time-like curves close to the 
massive object. Nevertheless, one can even argue that the restriction
imposed by such conjecture is more motivated by the 
limitations of our current knowledge of gravity rather than by 
true physical reasons. For example, the physical interpretation
of space-times with causality violating regions has been recently
investigated by a couple of authors in the framework of string
theories~\cite{horava, israel, drukker, gimon}. In these works,
it was found that such pathological regions go to a new phase
and a domain wall forms. Across the domain wall, the metric is
non-differentiable and the expected region with closed time-like
curves arises from the naive continuation of the metric
ignoring the domain wall. A similar mechanism could work for the
Tomimatsu-Sato space-times. A domain wall as a surface of the
astrophysical black hole candidates would be also consistent
with the non-observation of X-ray bursts from X-ray binaries 
with a black hole candidate: the accreting matter could be
converted to exotic stuff as soon as it hits the surface of the
compact object and no thermonuclear reaction would occur~\cite{abramo}.

The paper is organized as follows. In Sec.~\ref{s-ts2}, we review 
the basic properties of the TS2 space-time. In Sec.~\ref{s-motion}, 
we present our approach and, in Sec.~\ref{s-shadow}, we determine
the shape and the position of the shadow in TS2 and we compare
with the one of Kerr black holes. In Sec.~\ref{s-disc}, we
discuss our results and the possibility of detecting deviations
from the Kerr metric with future experiments. Summary and conclusions 
are reported in Sec.~\ref{s-con}. Throughout the paper we use natural 
units $G_N = c = 1$ and metric with signature $(-+++)$.

\section{The $\delta = 2$ Tomimatsu-Sato space-time \label{s-ts2}}

The Tomimatsu-Sato space-times form a family of stationary, 
axisymmetric, and
asymptotically flat exact solutions of the vacuum Einstein equations.
They are characterized by three parameters: the mass $M$, the spin 
$J$, and the deformation parameter $\delta$. The canonical form of
the line element of a stationary and axisymmetric space-time is
\be
ds^2 = - f \left(dt - \omega d\phi\right)^2 
+ \frac{1}{f}\left[e^{2\gamma}\left(d\rho^2 + dz^2\right)
+ \rho^2 d\phi^2\right] \, ,
\ee
where $f$, $\omega$, and $\gamma$ are functions of the quasi-cylindrical
coordinates $(\rho,z)$. The Tomimatsu-Sato solutions have
\be
f = \frac{A}{B} \, , \qquad
\omega = \frac{2 M q (1 - y^2) C}{A} \, , \qquad
e^{2\gamma} = \frac{A}{p^{2\delta} (x^2 - y^2)^{\delta^2}} \, .
\ee
$A$, $B$, and $C$ are polynomials, respectively of degree $2\delta^2$,
$2\delta^2$, and $2\delta^2 - 1$ in the prolate spheroidal coordinates
$(x,y)$, defined by
\be
\rho = \sigma \sqrt{(x^2 - 1)(1 - y^2)} \, , \qquad
z = \sigma x y \, .
\ee
The parameters $p$, $q$, and $\sigma$ are related to $M$, $J$, 
and $\delta$ as follows
\be
p^2 + q^2 = 1 \, , \qquad
q = \frac{J}{M^2} \, , \qquad
\sigma = \frac{M p}{\delta} \, .
\ee
The $\delta = 1$ solution is the Kerr space-time. For $\delta = 2$, 
$A$, $B$, and $C$ are given by 
\be
A &=& p^4 (x^2 - 1)^4 + q^4 (1 - y^2)^4 
- 2 p^2 q^2 (x^2 - 1)(1 - y^2) 
\left[2(x^2 - 1)^2 + 2(1 - y^2)^2 + 3(x^2 - 1)(1 - y^2)\right] \, , \\
B &=& \left[p^2 (x^2 + 1)(x^2 - 1) - q^2 (1 + y^2)(1 - y^2)
+ 2 p x (x^2 - 1)\right]^2
+ 4 q^2 y^2 \left[p x (x^2 - 1) + (p x + 1)(1 - y^2)\right]^2 \, , \\
C &=& - p^3 x (x^2 - 1)
\left[2 (x^2 + 1)(x^2 - 1) + (x^2 + 3)(1 - y^2)\right]
- p^2 (x^2 - 1)\left[4 x^2 (x^2 - 1) + (3 x^2 + 1)(1 - y^2)\right]
\nonumber \\ && + q^2 (p x + 1)(1 - y^2)^3 \, .
\ee

\begin{figure}
\par
\begin{center}
\includegraphics[height=6cm,angle=0]{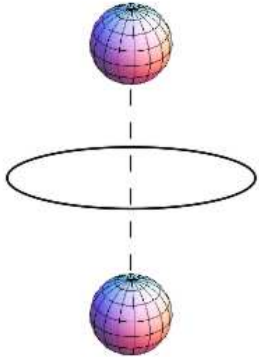}
\end{center}
\par
\vspace{-5mm} 
\caption{Structure of the $\delta = 2$ Tomimatsu-Sato space-time
in $(\rho,z)$ coordinates. There are two Killing horizons (here 
represented by two spherical 
surfaces), which are connected by the segment singularity 
(dashed line). On the equatorial plane, there is the ring
singularity (solid line).}
\label{f-pic}
\end{figure}

TS2 has quite peculiar properties. Here we just mention briefly
the main features, alerting the reader that there are numerous
mistakes, typos, and wrong sentences in the literature, and we
refer to~\cite{kodama} for more details. The structure of TS2 
is sketched in Fig.~\ref{f-pic}. On the equatorial plane,
there is a ring singularity where $B$ vanishes. As discussed
in~\cite{kodama}, the ring singularity has zero Komar
mass and an infinite circumference, in the sense that there
$g_{\phi\phi}$ diverges. For odd $\delta$, the segment I: 
$\rho = 0$, $|z| < \sigma$ (the surface $x = 1$) is an event horizon. 
In TS2, the space-time cannot be extended to the region $\rho < 0$. 
It is a segment singularity with a quite exotic structure, 
see~\cite{kodama}. It seems to carry the gravitational mass of 
the system. At $\rho = 0$ and $z = \pm \sigma$ ($x = 1$ and $y = \pm 1$), 
there are two Killing horizons, which are regular except at the point 
of connection with the segment singularity. Around the segment 
singularity and inside the ring singularity, $g_{\phi\phi} < 0$ 
and there are thus closed time-like curves. Such a causality 
violating region is the most suspicious property of TS2.
Using the standard definition of multipole moments for 
stationary and asymptotically flat space-times~\cite{hansen},
the quadrupole moment of the system is
\be
Q = \left(\frac{1}{4} + \frac{3}{4} \frac{J^2}{M^4}\right) M^3 \, ,
\ee
which is larger than the Kerr one $Q = J^2/M$ for $|q| < 1$.
Like all the Tomimatsu-Sato space-times, TS2 reduces to the
space-time of an extreme Kerr black hole in the special case 
$|q| = 1$.

\section{Motion of massless particles in TS2 \label{s-motion}}

In any stationary and axisymmetric space-time, there are three
conserved quantities: the mass of the test particle $\mu$, the
energy at infinity $E$, and the angular momentum at infinity 
along the symmetry axis, say $L_z$. This fact allows for the
separability with respect to the affine parameter $\lambda$
and the $t$ and $\phi$ coordinates of the Hamilton-Jacobi
equation. Since the Kerr space-time is of Petrov type D, there
is a fourth conserved quantity (the Carter constant $Q$) which
leads to the additional separability with respect to the other
two coordinates, at least in some coordinate systems. In general,
however, this last step is not possible and one has to 
integrate (numerically) the associated second order geodesic 
equations.

In order to compute the shadow of the massive object in
TS2, we use the following set-up. First, we adopt Schwarzschild 
coordinates $(r,\theta)$, which are related to the 
quasi-cylindrical coordinates $(\rho,z)$ by
\be\label{sch-coo}
\rho = \sqrt{r^2 - 2 M r + q^2 M^2} \sin\theta \, , \qquad
z = (r - M) \cos\theta \, .
\ee
Then, we consider an observer at infinity with inclination 
angle $i$. In the numerical integration, the radial coordinate
of the observer is $r = 10^5~M$: at such a distance, the 
space-time can be considered flat. The image plane of the
observer is the plane orthogonal to his line of sight, see
e.g.~\cite{bardeen, chandra}. The plane is provided by a system 
of cartesian coordinates $(X,Y)$. 
We consider the photons on the image plane
with 3-momentum perpendicular to the plane; that is, with 
${\bf k} = k \hat{\bf Z}$, where $\hat{\bf Z}$ is the unit
vector orthogonal to the image plane. 
Writing the position and the momentum of the photon in the 
prolate spheroidal coordinates, we solve the geodesic 
equation with a fourth order Runge-Kutta-Nystrom method.

Some photons approach the object and then come back to infinity 
(non-captured orbits). Other photons are instead captured by 
the massive object (captured orbits): in the case of black hole, 
they would cross the event horizon, while in TS2 
they can either cross one of the horizons at $\rho = 0$ and 
$z = \pm \sigma$, or end at the segment or the ring singularities.
In any case, these photons cannot come back to infinity.
The shadow is the image formed on the image plane by the set 
of photons captured by the massive object.
We have corotating orbits, when the photon angular momentum
is parallel to the spin of the massive object, and
counterrotating orbits, in the opposite case. The minimum 
radius from the object of the non-captured orbits is
larger than the photon capture radius, which is the radius
of the innermost photon orbit. In Table~\ref{t-ts},
we reported the photon capture radius for corotating and 
counterrotating orbits on the equatorial plane, for a few
values of the parameter $q$. In Table~\ref{t-kerr}, there
are the same quantities for the Kerr case.
In Tables~\ref{t-ts} and \ref{t-kerr}, we show also the
impact parameter, $b = \sqrt{X^2 + Y^2}$, of the orbits
separating non-captured and captured photons.

\begin{table}
\begin{center}
\begin{tabular}{|c||c|c|c||c|c|c|}
\hline
$\quad |q| \quad$ & $\quad x_{\gamma}^{+} \quad$ & $\quad r_{\gamma}^{+} \quad$ & $\quad b^+ \quad$ & $\quad x_{\gamma}^{-} \quad$ & $\quad r_{\gamma}^{-} \quad$ & $\quad b^- \quad$\\
\hline
\hline
0.0 & 4.01 & 3.17 & 5.38 &  4.01 & 3.17 & 5.38 \\ \hline
0.1 & 3.81 & 3.08 & 5.19 &  4.26 & 3.29 & 5.56 \\ \hline
0.3 & 3.45 & 2.84 & 4.79 &  4.87 & 3.47 & 5.91 \\ \hline
0.5 & 3.17 & 2.56 & 4.34 &  5.86 & 3.65 & 6.24 \\ \hline
0.7 & 2.95 & 2.22 & 3.80 &  7.66 & 3.80 & 6.55 \\ \hline
0.9 & 2.82 & 1.72 & 3.07 & 13.39 & 3.94 & 6.85 \\
\hline
\end{tabular}
\end{center}
\caption{Photon capture radius in prolate spheroidal and 
Schwarzschild coordinates on the equatorial plane for 
corotating ($+$) and counterrotating ($-$) orbits in TS2 for
a few values of $|q|$. $b$ is the impact parameter.
$x$, $r$, and $b$ are in units $M=1$.}
\label{t-ts}
\end{table}

\begin{table}
\begin{center}
\begin{tabular}{|c||c|c||c|c|}
\hline
$\quad |q| \quad$ & $\quad r_{\gamma}^{+} \quad$ & $\quad b^+ \quad$ & $\quad r_{\gamma}^{-} \quad$ & $\quad b^- \quad$ \\
\hline
\hline
0.0 & 3.00 & 5.20 & 3.00 & 5.20 \\ \hline
0.1 & 2.88 & 4.99 & 3.11 & 5.39 \\ \hline
0.3 & 2.63 & 4.56 & 3.33 & 5.77 \\ \hline
0.5 & 2.35 & 4.10 & 3.53 & 6.13 \\ \hline
0.7 & 2.01 & 3.57 & 3.73 & 6.49 \\ \hline
0.9 & 1.56 & 2.93 & 3.91 & 6.83 \\
\hline
\end{tabular}
\end{center}
\caption{As in Table~\ref{t-ts}, in the case of Kerr space-time.}
\label{t-kerr}
\end{table}

\section{Prediction of the shadow in TS2\label{s-shadow}}

The shadow of a black hole is the dark area over a bright
background viewed by a distant observer when the black hole
is in front of a planar light source~\cite{bardeen}. In a
realistic case, the emission occurs from the accreting gas
surrounding the black hole, but one still observes the same
dark shape~\cite{melia, rohta}. The shadow can also be 
determined by studying the photon orbits leaving the
image plane of the distant observer and intersecting the
horizon.

In TS2, the shadow is formed by 
all the photons that leave the image plane and are captured
by the gravitational field of the massive object. In the 
reference frame of the observer, the photon initial conditions 
are given by $(X,Y,Z=0)$ and $(0,0,k)$. In order to compute
the shadow, we write the photon initial conditions in the 
prolate spheroidal coordinates. Then, we integrate numerically 
the standard second order geodesic equations.
Like in the Kerr case, the shape and the position of the 
shadow depend only on the observer inclination angle, $i$,
and the dimensionless spin parameter of the object, $q$.

\begin{figure}
\par
\begin{center}
\includegraphics[height=7cm,angle=0]{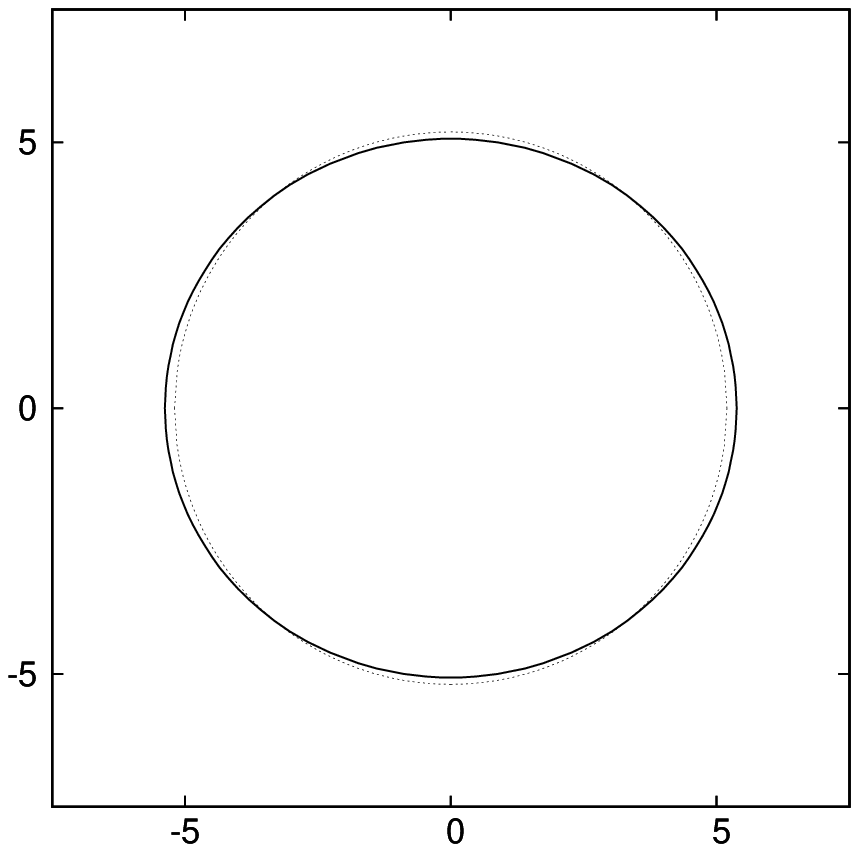} \hspace{-2.5cm}\includegraphics[height=7cm,angle=0]{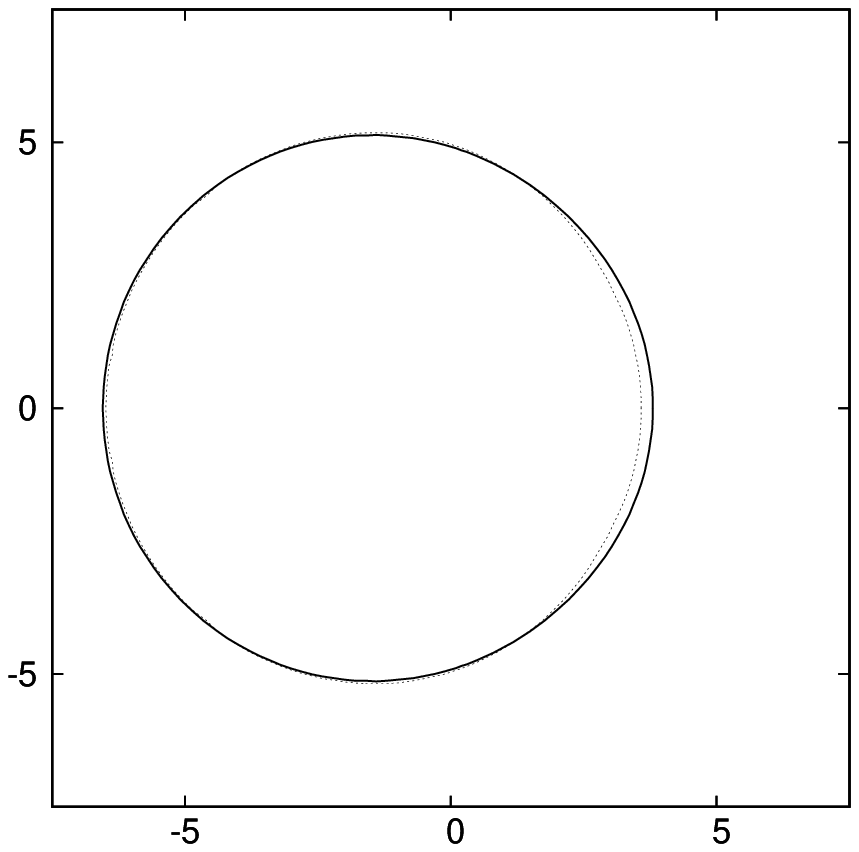} \\ \vspace{.3cm}
\includegraphics[height=7cm,angle=0]{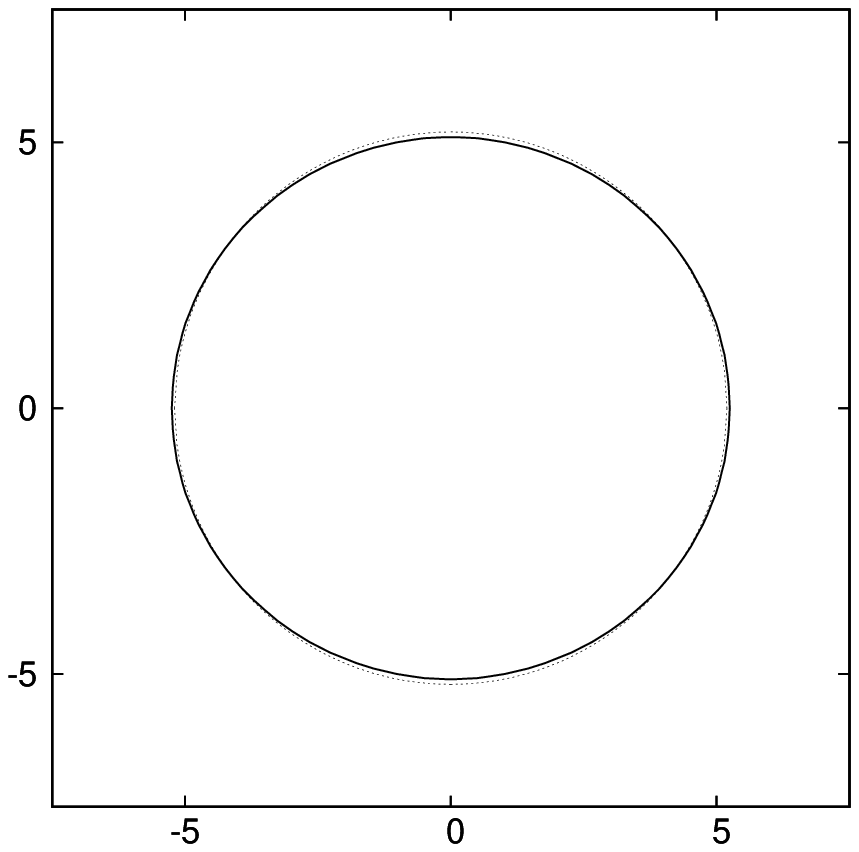} \hspace{-2.5cm}
\includegraphics[height=7cm,angle=0]{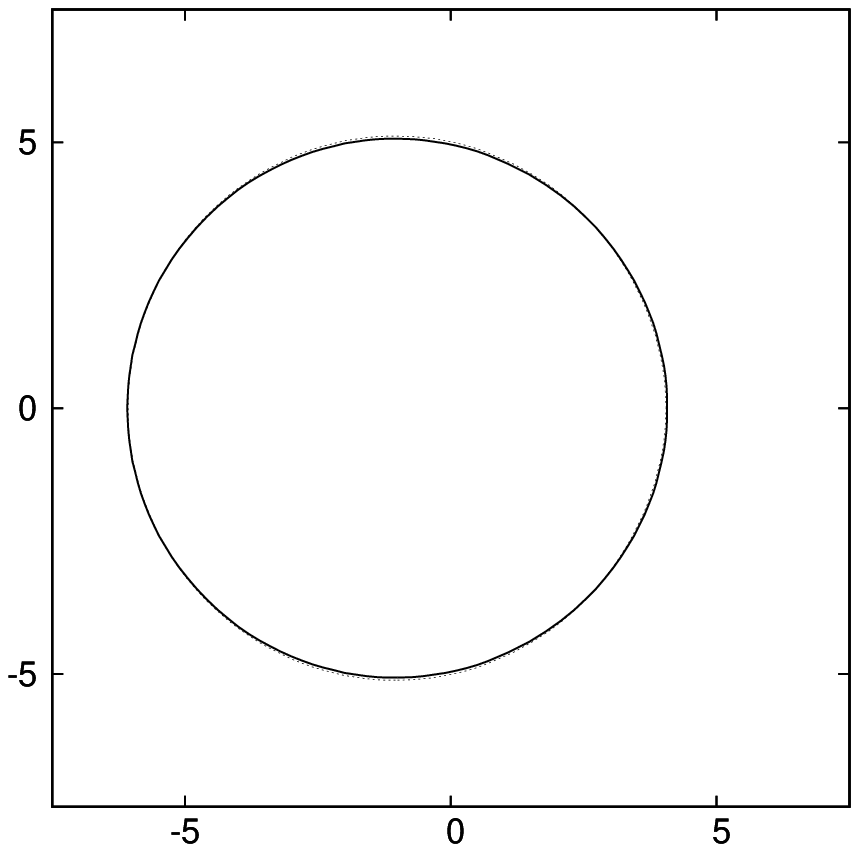} \\ \vspace{.3cm}
\includegraphics[height=7cm,angle=0]{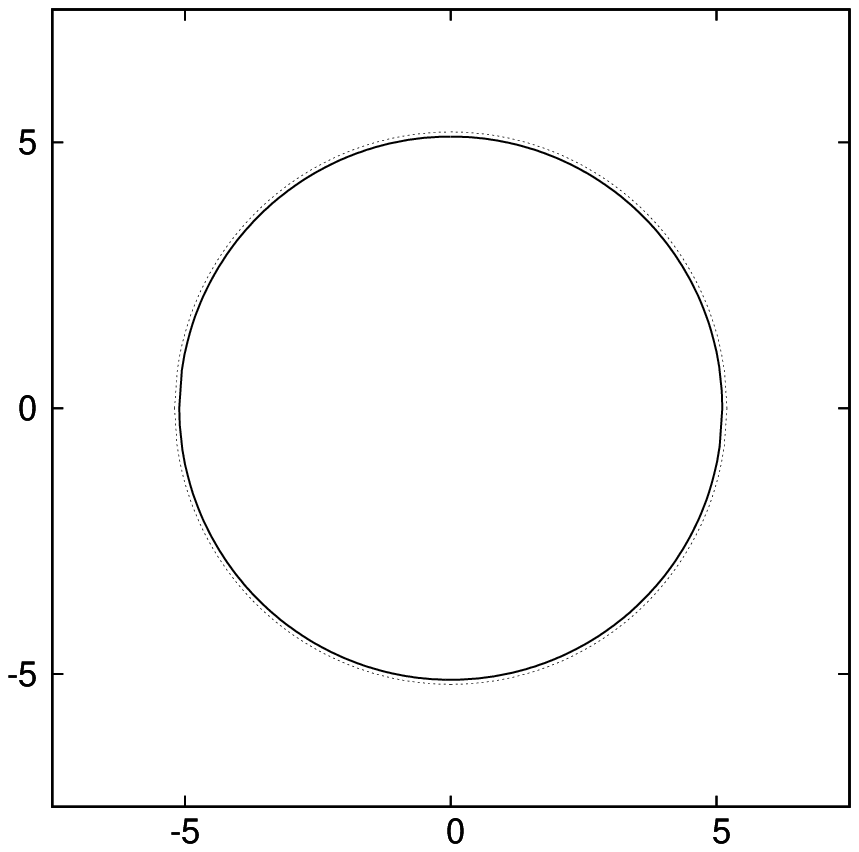} \hspace{-2.5cm}
\includegraphics[height=7cm,angle=0]{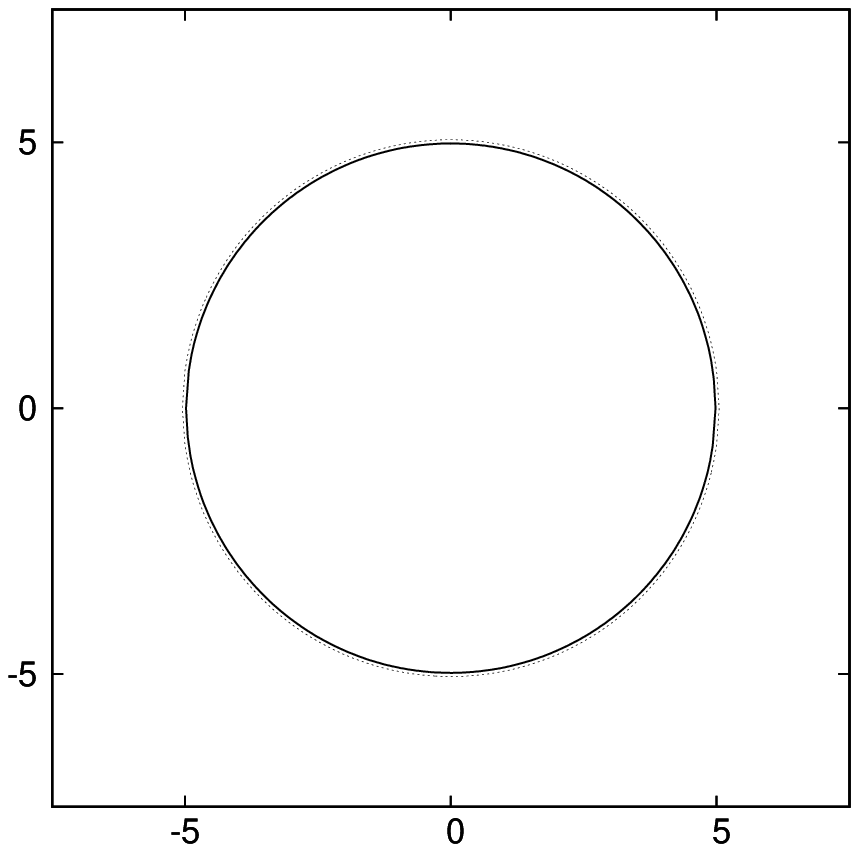} 
\end{center}
\par
\vspace{-5mm} 
\caption{Contour of the shadow in $\delta = 2$
Tomimatsu-Sato space-time (solid lines) and in Kerr 
space-time (dotted lines) in the $XY$-plane of the
distant observer, for spin parameter $q = 0$
(left column) and $q = 0.7$ (right column). The viewing
angle is respectively $i = 90^\circ$ (top panels), 
$45^\circ$ (central panels), and $0^\circ$ (bottom panels).
Horizontal and vertical axis in units $M = 1$.}
\label{f-main}
\end{figure}

\begin{figure}
\par
\begin{center}
\includegraphics[height=6cm,angle=0]{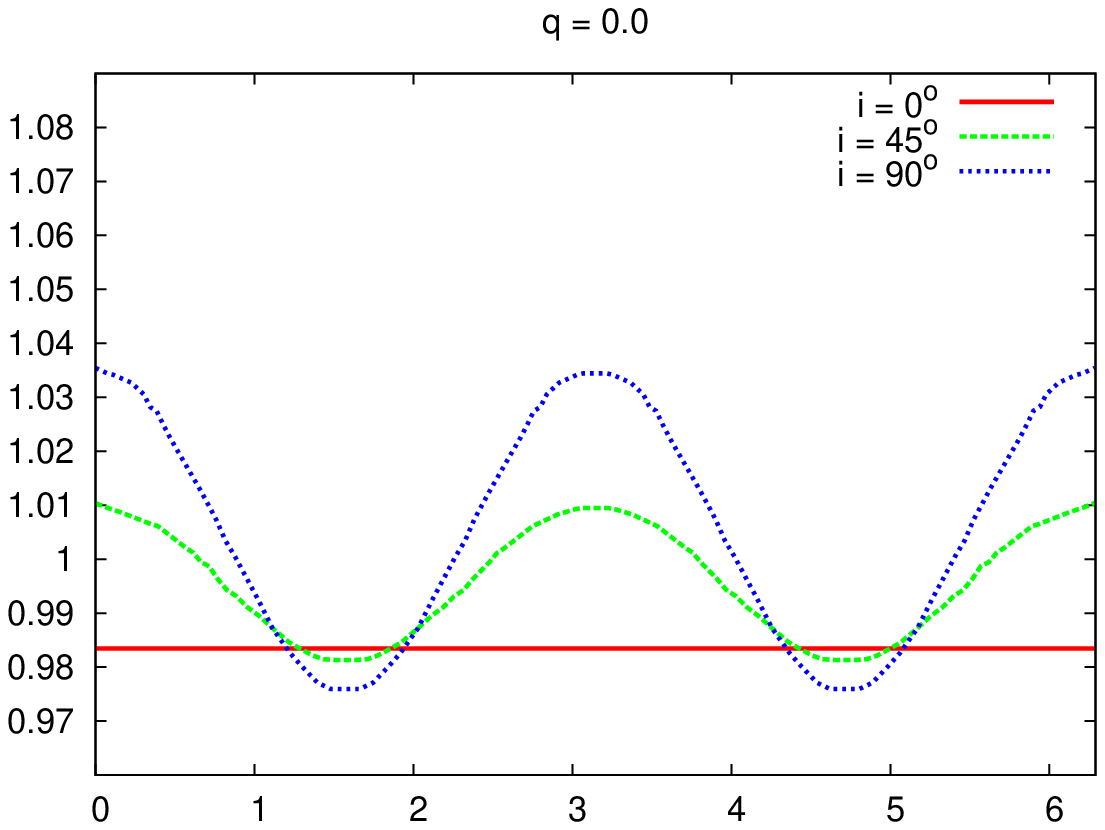} \hspace{-0.5cm}
\includegraphics[height=6cm,angle=0]{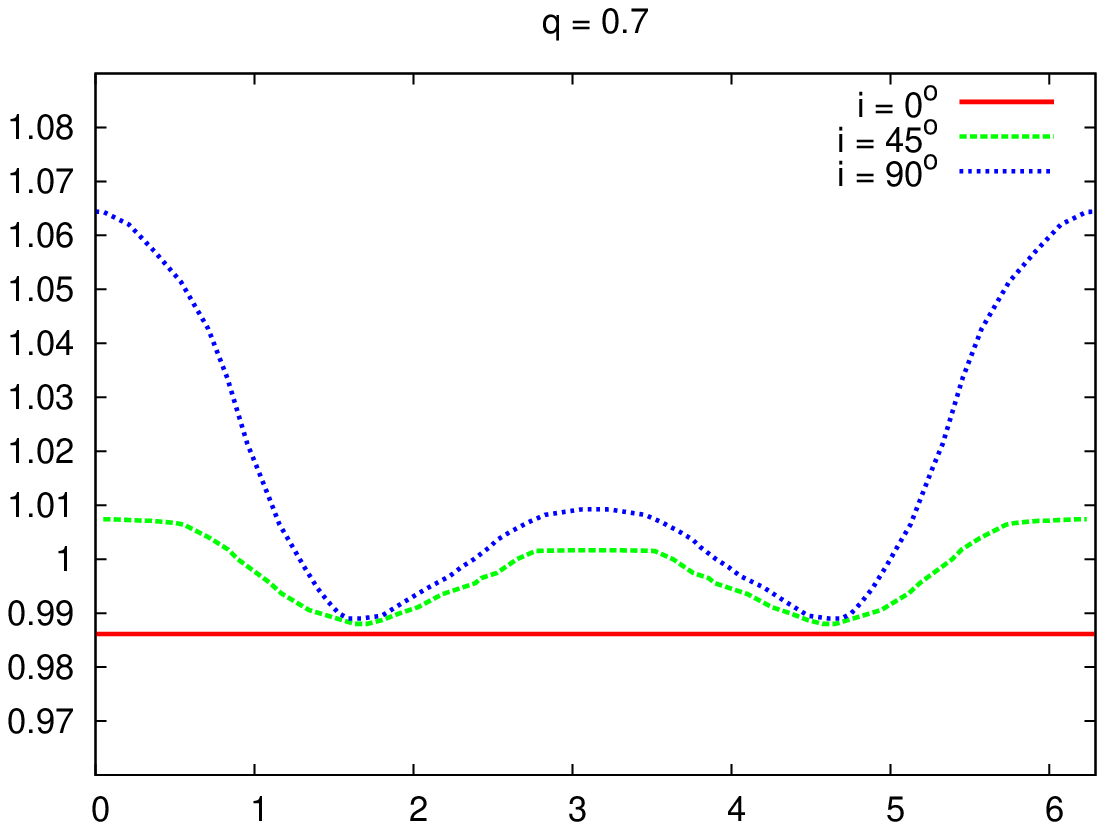} 
\end{center}
\par
\vspace{-5mm} 
\caption{$R_{TS2}/R_{Kerr}$ as a function of the polar
coordinate $\Theta$ for the shadows shown in Fig.~\ref{f-main}.
Left panel: spin parameter $q = 0$. Right panel: spin parameter 
q = 0.7.}
\label{f-ratio}
\end{figure}

The results of our study is summarized in Fig.~\ref{f-main},
where we show the contour of the shadows for 
$q = 0, \, 0.7$ and $i = 90^\circ$, $45^\circ$, and $0^\circ$.
The horizontal and vertical axis are respectively the
$X$ and $Y$ axis of the image plane of the distant observer.
The asymmetry of the shadow with respect to the $Y$ axes is
due to the difference of corotating and counterrotating 
orbits and is present either in Kerr and in TS2.
For an observer on the equatorial plane ($i = 90^\circ$),
the shadow in TS2 is a bit flatter than the Kerr
case. The difference is however small: for $|q| \lesssim 0.9$, 
the size of the shadow changes by about 3 -- 4\%, while, for 
fast-rotating object, the difference is smaller, and becomes 
negligible when $|q|$ is very close to 1 (in the special case
$|q| = 1$, TS2 reduces to an extreme Kerr black hole). 
For smaller and larger inclination angles, the two shadows
are more and more similar and, for $i = 0^\circ$ or $180^\circ$,
in both cases we find a circle, as one can easily argue 
from the axial symmetry of the space-times. In TS2, the circle
is only slighter smaller than the one in Kerr space-time. For 
moderate values of the spin parameter, the radius of the two 
circles differs by about 2\%. 
The contour of the shadow on the image plane of the distant
observer can also be described by the polar coordinates $(R,\Theta)$.
They are related to the cartesian coordinates $(X,Y)$ by 
\be
X = R \cos\Theta \, , \qquad Y = R \sin\Theta \, .
\ee 
The contour of every shadow can thus be written as $R = R(\Theta)$.
In Fig.~\ref{f-ratio}, we plot the quantity $R_{TS2}/R_{Kerr}$
as a function of $\Theta$, for the cases shown in Fig.~\ref{f-main}.

\section{Discussion \label{s-disc}}

The shadows in the Kerr and TS2 space-times are surprisingly 
similar. In particular, we do not see any feature associated to 
the ring singularity of TS2. The reason is that the shadow is 
determined by the photon capture radius, which is always larger 
than the radius of the ring singularity. In particular, for 
$q = 0$ the ring singularity is on the singular surface $x = 1$, while the 
photon capture radius is at $x_\gamma \approx 4$ (see Table~\ref{t-ts}). 
For $q = 0.7$, the
ring singularity is at $x \approx 1.09$; here the photon capture
radius is at $x_\gamma \approx 2.95$. Some confusion may be also 
generated by the use of the prolate spheroidal coordinates $(x,y)$,
which allows for writing the metric in analytic form.
This point can be understood better if we consider the limit 
$|q| \rar 1$, in which TS2 reduces to an extreme Kerr black hole. 
In the prolate spheroidal coordinates, the radius of the ring 
singularity increases as $|q|$ increases, and goes to infinity 
when $|q| = 1$. In the Schwarzschild coordinates, the radius of 
the ring singularity decreases as $|q|$ increases, and is 
``absorbed'' by the event horizon when $|q| = 1$.

The ring singularity has several unphysical features. In particular,
it is a curvature singularity and marks the boundary of the 
causality violating region. We can thus expect that new physics,
if any, appears before reaching the ring singularity and that 
here the space-time is quite different from the one predicted by 
the TS2 metric. Nevertheless, one may also consider an exact TS2
space-time and wonder how the ring singularity appears to a
distant observer if, for example, it is a bright source of photons.
The distant observer can see multiple images, in the sense that
a single point of the ring singularity produces a brighter primary
image and less bright secondary images. The
primary image is associated with the shortest photon orbit
connecting the point of the ring singularity with the image plane.
The other images are associated with photon orbits turning around
the massive objects before reaching the distant observer. On the 
basis of the symmetries of the space-time, one can immediately say
that an observer on the equatorial plane sees the ring singularity
as a segment at $Y = 0$ with the two ends at the boundary of the
shadow. An observer with $i = 0^\circ$ or $180^\circ$ sees instead
a set of circles inside the shadow. The primary image is the
circle with smaller radius. For $q = 0$, the radius of the
primary image on the image plane of the observer is about 2.8~$M$. 
For rotating objects, this radius is a bit smaller: for example,
when $q = 0.7$, the radius is about 2.6~$M$. For different
inclination angles, the observer sees ellipses or arcs.

The key ingredient to test new physics 
is the accurate determination of the shape, rather than of the 
absolute size, of the shadow. The size is indeed determined by 
the mass and the distance of
the massive objects, which can be unlikely measured with good
precision. This means, in particular, that for viewing angles 
close to $0^\circ$ or $180^\circ$ it is impossible to distinguish
the two metrics, since in both cases we have a circle with a 
radius which is slightly different (see Fig.~\ref{f-main}).
In the non-rotating case and for a viewing angle $i = 90^\circ$, 
the ratio between the larger and smaller axes of the shadow in
TS2 is about 1.06. So, if we consider the supermassive black hole 
candidate at the center of the Galaxy, SgrA*, whose shadow should
have a diameter
$\sim 50$~$\mu$as, we need at least a resolution of a few $\mu$as
in order to rule out the TS2 metric. For the supermassive black hole 
candidate in M87, the angular size of its shadow should be
about 20~$\mu$as and we would thus need a resolution around 1~$\mu$as.
Unfortunately, near future experiments on the Earth or space 
missions such as VSOP-2~\cite{web1, web2} will not be able to 
distinguish the Kerr metric from the TS2 one. Indeed, we already 
know that such experiments will not be able to measure the spin of 
a black hole~\cite{melia, rohta}, assuming the Kerr metric, where 
the size of the shadow along its axis of symmetry changes 
even by about 14\% (for $i = 90^\circ$) between a Schwarzschild 
and a maximally rotating black hole, while the 
one perpendicular to its axis of symmetry is very similar. 
However, more advanced space missions may distinguish the two cases. 
Sub-mm space interferometers (e.g. VSOP-3) will be capable of 
measuring the shape of the shadow of nearby super-massive black 
hole candidates at the level of 1\%, which may be enough to distinguish 
the Kerr from the TS2 case, at least for $|q| \lesssim 0.9$ 
and for a viewing angle $\sim 90^\circ$. In less favorable
circumstances, more advanced experiments, such as X-ray
interferometers, might be necessary.

TS2 is just a special solution in the Tomimatsu-Sato family, with 
no particular properties except the fact it can be obtained as the 
superposition of two Kerr black holes.  
It is worth mentioning what happens for 
higher values of the deformation parameter $\delta$. The expression 
of the metric becomes more and more complicated as $\delta$ increases. 
Nevertheless, we can argue that the shadow becomes only slightly 
more flattened than the case $\delta = 2$. Indeed, 
the shadow, being determined by the photon orbits around the 
massive object, can be seen as a measurement of the multipole 
moments associated to the space-time. After the mass $M$ and the 
spin $J$, the most important term is the quadrupole moment $Q$,
which is given by
\be
Q = \left( q^2 + \frac{\delta^2 - 1}{3 \delta^2} p^2 \right) M^3 \, .
\ee
In the simplest case of $q = 0$, the quadrupole moment is
$Q = 0$ for $\delta = 1$, $Q = M^3/4$ for $\delta = 2$, and approaches
the limit $Q = M^3/3$ as $\delta$ increases. So, if an observer on
the equatorial plane see a shadow whose symmetry axes differ by
about 6\% for $\delta = 2$, the difference becomes only a bit larger,
approximately 8\%, when the value of the deformation parameter is very 
large.

\section{Conclusions \label{s-con}}

Within the next decade, very long baseline interferometry techniques 
will likely be able to image the surrounding environment of some 
super-massive black hole candidates, with resolution at the level 
of the black hole event horizon. These experiments will open the 
possibility of studying strong gravity and the actual nature of 
the final product of the gravitational collapse. One of the main 
goal is the observation of the shadow, whose shape and position 
directly depend on the metric of the space-time around the massive 
object.

In this paper, we have investigated the shape and the position of 
the shadow which would be observed in a $\delta = 2$ Tomimatsu-Sato
space-time, and compared with the one expected in the Kerr case.
The $\delta = 2$ Tomimatsu-Sato space-time is a stationary,
axisymmetric, and
asymptotically flat exact solution of the vacuum Einstein equations
which violates the Cosmic Censorship Conjecture. It is not a
black hole solution. Despite several significant differences between
the two space-times, the associated shadows are quite similar.
In the $\delta = 2$ Tomimatsu-Sato space-time, the shadow is more
flattened and an accurate measurement of its shape, at the level 
of $\sim 1\%$, may observe deviations from the Kerr metric, at
least for certain values of the spin parameter $q$ and the
viewing angle $i$. So, near future ground based experiments or
space missions like VSOP-2 will not be able to detect
any difference. With more advanced space missions in the future, it is
still challenging, but not out of reach.

However, our conclusion that we have to wait for 15-20 years
may be too pessimistic. Since around naked singularities the 
gravitational force can be repulsive~\cite{sim1}, outflows of 
gas can be produced~\cite{sim2}. Similar phenomena are among the
science goals of experiments like VSOP-2.

\begin{acknowledgments}
We would like to thank Hideo Kodama and Rohta Takahashi 
for useful discussions and suggestions.
The work of C.B. was partly supported by the JSPS 
Grant-in-Aid for Young Scientists (B) No. 22740147.
This work was supported by World Premier International 
Research Center Initiative (WPI Initiative), MEXT, Japan.
\end{acknowledgments}


\end{document}